\documentclass[aps,prl,twocolumn,groupedaddress]{revtex4-1}
\usepackage{graphicx}
\usepackage{dcolumn}
\usepackage{bm}
\usepackage{amssymb}
\usepackage{color}

\begin{document}


\title{Shape transition from elliptical to cylindrical membrane tubes induced by chiral crescent-shaped protein rods}

\author{Hiroshi Noguchi}
\email[]{noguchi@issp.u-tokyo.ac.jp}
\affiliation{
Institute for Solid State Physics, University of Tokyo,
 Kashiwa, Chiba 277-8581, Japan}


\begin{abstract}
Proteins often form chiral assembly structures on a biomembrane.
However, the role of the chirality in the interaction with an achiral membrane is poorly understood.
Here, the differential behavior between chiral and achiral crescent-shaped protein rods was investigated 
using meshless membrane simulations.
The achiral rods deformed the membrane tube into an elliptical shape by stabilizing the edges of the ellipse.
In contrast, the chiral rods formed a helical assembly that generated a cylindrical membrane tube with a constant radius
in addition to the elliptical tube.
This helical assembly could be further stabilized by the direct side-to-side attraction between the protein rods.
These results agree with experimental findings of the constant radius of membrane tubules induced by 
the Bin/Amphiphysin/Rvs (BAR) superfamily proteins.
\end{abstract}



\maketitle

The molecular chirality and single-handedness have been under strong selection in evolution~\cite{blac10}.
Proteins consist of L-amino acids and form chiral structures
from a local secondary structure, the right-handed $\alpha$-helix, to micrometer-scale assemblies such as a microtubule.
This chirality is important for the recognition of molecular binding, and the
interactions between chiral molecules are often governed by a geometrical match.
The molecular chirality is crucial for the binding affinity to DNA double helices~\cite{corr07}.
For example, a right-handed molecule strongly binds to the minor groove of the DNA helix, whereas binding of the left-handed form 
may be prevented due to excluded volume interactions.
However, 
the role of chirality  in the interaction between chiral and achiral structures
remains poorly understood in comparison with the interactions between chiral structures.
Accordingly, the aim of this study is to clarify the mechanisms underlying the chirality of membrane-binding proteins 
in the interaction with deformable achiral membranes.

In living cells, lipid membranes are maintained in a fluid phase,
in which the chiral molecular interactions between chiral lipids are smeared out
so that the chirality does not appear in membrane dynamics.
However, in a gel phase, chiral amphiphiles can form helical-ribbon structures, which
transform to fluid vesicles at high temperature \cite{naka85,schn93,shim05}.
Here, we consider a biomembrane as a two-dimensionally isotropic achiral fluid membrane.

Many proteins are known to bind biomembranes and consequently reshape them~\cite{mcma05,shib09,baum11,mcma11,suet14,joha15}.
Such proteins are often found in a helical assembly formation. 
For example, dynamins assemble on the neck of a clathrin-coated membrane bud 
and form a helix surrounding the neck to induce the membrane fission~\cite{anto16}.
Proteins containing a Bin/Amphiphysin/Rvs (BAR) domain, which consists of a banana-shaped dimer,
bind the membrane and bend it along the BAR domain axis via scaffolding~\cite{itoh06,masu10,zhao11,mim12a,simu15a}.
The BAR domains have chirality, and helical alignment on the membrane tubes has been observed by electron microscopy~\cite{mim12a,fros08,adam15}.
However, the role of this helicity is not known. 
At low concentration, BAR-containing proteins solely
 bind the membrane with maintained ability as membrane scaffolding~\cite{simu16}.
The two-dimensional crystal alignment of the BAR domains induces substantially greater bending ability;
however, it is unclear whether a helical structure is essential for this effect, 
and if achiral crystal assembly might yield the same degree of bending.
Since it is nearly impossible to separate the chirality and regularity of a protein assembly experimentally, 
here, the general effects of the chirality were investigated using numerical simulations in which
 chirality can be readily switched on and off.
The attractive interactions between specific sites of the BAR domains were then considered in terms of elucidating 
the origin of the regular assembly~\cite{fros08,adam15,itoh16}.
The I-BAR protein, Ivy1p, forms a filament by side-to-side attractions, even in the absence of membranes~\cite{itoh16}.
Therefore, the effects of the side-to-side attraction of chiral proteins on membrane tubulation were also investigated.

The binding of BAR proteins to the membranes and resulting shape deformation
have previously been simulated using various approaches from atomistic molecular dynamics to mesoscale coarse-grained models \cite{bloo06,arkh08,yu13,simu13,simu15,take17,rama13,nogu14,nogu15b,nogu16,nogu16a,nogu17,nogu17a}.
Although the atomistic and coarse-grained molecular models of the BAR domains~\cite{bloo06,arkh08,yu13,simu13,simu15,take17}
have chirality, the chirality effects have not been investigated to date.

In this study, we examined the effects of such chirality in crescent-shaped protein rods
on the shape deformation of membrane tubes using coarse-grained membrane simulations.
A tubular membrane is a well-developed experimental setup~\cite{baum11};
the tubular tether membrane is extended from a vesicle by optical tweezers 
so that the tube radius is controllable by manipulating the mechanical force.
We examined how the rod chirality changes the assembly structure and shapes of the membrane tubes.
The chiral rods behaved similarly to the achiral rods at a low curvature $C_{\rm {rod}}$ but
formed a helical cylinder at a high curvature. This circular tube could not be induced by the achiral rods.
In addition, the side-to-side attraction between the chiral rods reinforces this cylinder formation.

\begin{figure}
\includegraphics[width=7.cm]{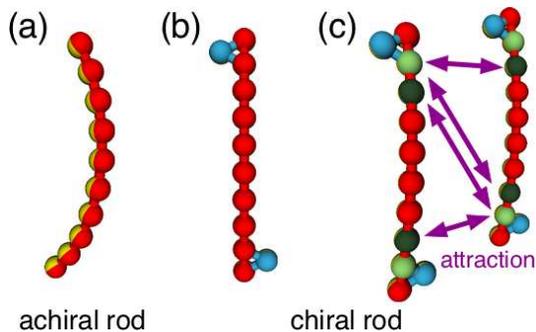}
\caption{
Protein models. (a) Achiral crescent-shaped rod. (b) Chiral crescent-shaped rod. 
(c) Chiral crescent-shaped rod with attractive interaction
between the second and third segments from both rod ends.
The orientation vector ${\bf u}_i$ lies along the direction from the 
light gray (yellow) to dark-gray (red, green, or blue) hemispheres.
}
\label{fig:model}
\end{figure}

The fluid membrane is modeled as single-layer self-assembled particles,
in which the membrane properties are highly tunable~\cite{nogu09,nogu06,shib11}.
A BAR protein is assumed to be strongly adsorbed onto the membrane; thus
the BAR and the membrane beneath it are modeled together 
as a chiral or achiral crescent-shaped rod with the spontaneous curvature of $C_{\rm {rod}}$ along the rod axis.
To construct a minimum model of chiral protein rods,
our previously reported linear protein model~\cite{nogu14,nogu15b,nogu16,nogu16a,nogu17,nogu17a} is modified 
by adding two particles
in the right-handed positions [Figs.~\ref{fig:model}(a) and (b)].
The details of the meshless membrane model and achiral protein rods are described 
in Ref.~\citenum{shib11} and Refs.~\citenum{nogu14,nogu16a}, respectively.

The membrane tube consists of $N$ particles.
The position and orientation vectors of the $i$-th particle are ${\bm{r}}_{i}$ and ${\bm{u}}_i$, respectively.
The membrane particles interact with each other via a potential $U=U_{\rm {rep}}+U_{\rm {att}}+U_{\rm {bend}}+U_{\rm {tilt}}$,
in which $U_{\rm {rep}}$ represents an excluded volume interaction with diameter $\sigma$,
$U_{\rm {att}}$ is the attractive potential to implicitly account for 
the effects of the solvent, and $U_{\rm {bend}}$ and $U_{\rm {tilt}}$ are
the bending and tilt potentials given by 
 $U_{\rm {bend}}/k_{\rm B}T=(k_{\rm {bend}}/2) \sum_{i<j} ({\bm{u}}_{i} - {\bm{u}}_{j} - C_{\rm {bd}} \hat{\bm{r}}_{i,j} )^2 w_{\rm {cv}}(r_{i,j})$
and $U_{\rm {tilt}}/k_{\rm B}T=(k_{\rm{tilt}}/2) \sum_{i<j} [ ( {\bm{u}}_{i}\cdot \hat{\bm{r}}_{i,j})^2
 + ({\bm{u}}_{j}\cdot \hat{\bm{r}}_{i,j})^2  ] w_{\rm {cv}}(r_{i,j})$,  respectively,
where ${\bm{r}}_{i,j}={\bm{r}}_{i}-{\bm{r}}_j$, $r_{i,j}=|{\bm{r}}_{i,j}|$,
 $\hat{\bm{r}}_{i,j}={\bm{r}}_{i,j}/r_{i,j}$, $w_{\rm {cv}}(r_{i,j})$ is a weight function,
and  $k_{\rm B}T$ denotes the thermal energy.
The spontaneous curvature $C_0$ of the membrane is 
given by $C_0\sigma= C_{\rm {bd}}/2$~\cite{shib11}.
For this study, the parameters $C_0=0$ and $k_{\rm {bend}}=k_{\rm{tilt}}=10$ were adopted in all cases 
except for the membrane particles belonging to the protein rods.
The membrane was given mechanical properties that are typical of lipid membranes:
a bending rigidity $\kappa/k_{\rm B}T=15 \pm 1$,
area of the tensionless membrane per particle $a_0/\sigma^2=1.2778\pm 0.0002$, and
area compression modulus $K_A\sigma^2/k_{\rm B}T=83.1 \pm 0.4$.
Moreover, the edge line tension $\Gamma\sigma/k_{\rm B}T= 5.73 \pm 0.04$ is set to be sufficiently large to prevent membrane rupture 
in the present simulations.

For modeling the achiral protein rods, $10$ membrane particles are linearly connected by the harmonic bond and bending potentials
 (the rod length $r_{\rm {rod}}=10\sigma$).
A relatively stronger bending rigidity, $k_{\rm {bend}}=k_{\rm{tilt}}=80$ is used for the protein rods than the membrane,
since the protein binding stiffens the membrane.
In the chiral protein rod, the particles between the first and second particles of both rod ends 
are  right-handedly added in a hook formation, as shown in Fig.~\ref{fig:model}(b),
and the excluded volumes of these two particles generate the chiral interactions between the rods.
For three harmonic bonds to form the triangle including the hock particle, three times greater bond coefficient 
is used to present a flip of the hock particle to the opposite ({\it i.e.}, left-handed) site.

Short and long membrane tubes with $N=4800$ and $9600$ are simulated
with a constant low protein density in which $40$ and $80$ protein rods are embedded, respectively.
Two average radii of the membrane tube $R_{\rm cyl}/r_{\rm rod}=1.18$ and $1.31$ 
corresponding to tube length $L_z/r_{\rm rod}= 0.00167N$ and $0.0015N$ are employed, respectively,
for both $N=4800$ and $9600$. 
The curvature of the protein rods is varied from $C_{\rm {rod}}r_{\rm {rod}}=0$ to $3.5$,
and zero spontaneous side curvatures~\cite{nogu16} in perpendicular to the rod axis is adopted.
These rod length and curvatures are within the typical range of known BAR proteins.
The BAR domain length ranges from $13$ to $27$ nm \cite{masu10}
and the rod curvatures are varied from negative to positive values.
Among the BAR proteins, APPL1 BAR-PH has the maximum curvature reported with  
a radius of the curvature is $5.5$ nm and the length is $17$ nm, 
{\it i.e.}, $C_{\rm {rod}}r_{\rm {rod}} \simeq 3$~\cite{zhu07}.
Molecular dynamics with a Langevin thermostat is employed~\cite{shib11,nogu11}.
In addition to canonical ensemble simulations,
replica exchange molecular dynamics (REMD) \cite{huku96,okam04} for the rod curvature $C_{\rm {rod}}$ \cite{nogu14} 
is used to accurately obtain the thermal equilibrium states at $N=4800$.
The error bars are estimated from three or more independent runs.

\begin{figure}
\includegraphics{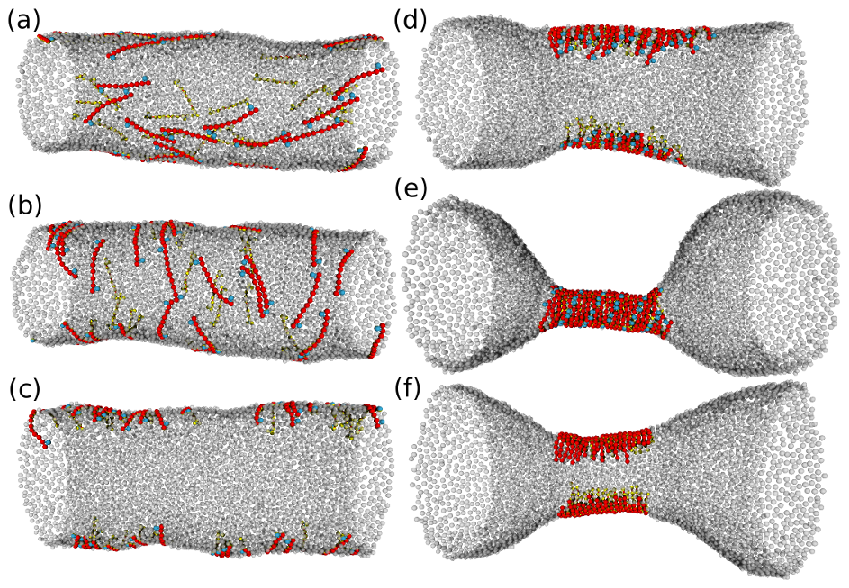}
\includegraphics{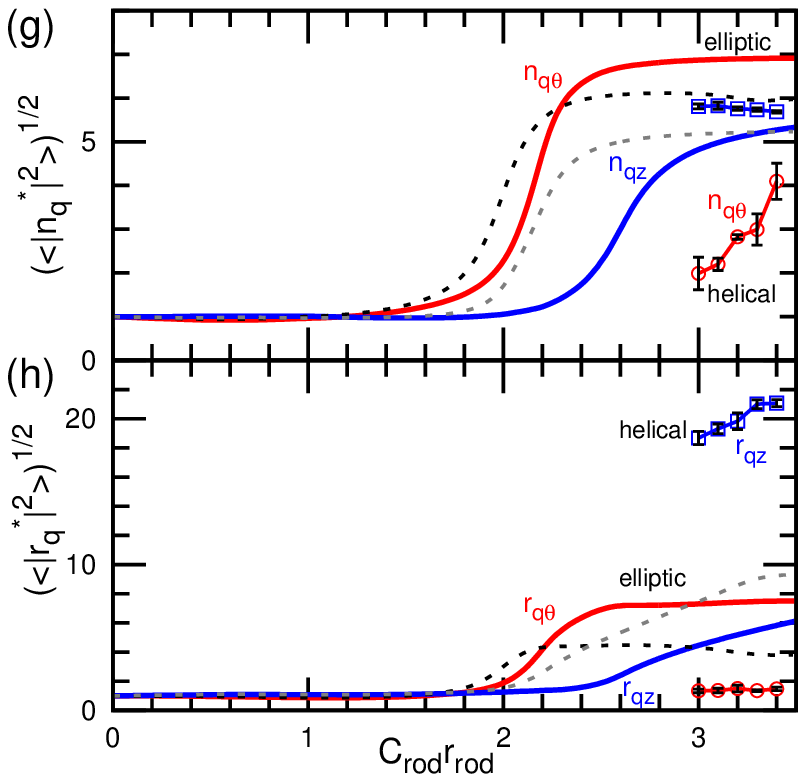}
\caption{
Membrane tube deformation induced by chiral and achiral protein rods 
without direct attraction between the rods at $R_{\rm cyl}/r_{\rm rod}=1.31$ and $N=4800$.
(a)--(f) Snapshots at (a) $C_{\rm {rod}}r_{\rm {rod}}=0$, (b) $1.5$, 
and (c) $2.3$, and (d), (e) $3.3$ for the chiral rods; and (f) $C_{\rm {rod}}r_{\rm {rod}}=3.3$ for the achiral rods.
The membrane particles are displayed as transparent gray spheres.
(g),(h) Fourier amplitudes of (g) rod densities and (h) membrane shapes.
The solid and dashed lines represent the data for REMD of the chiral and achiral protein rods, respectively.
The circles and squares with solid lines represent the $q\theta$ and $qz$ modes, respectively, 
for the canonical simulations of
the helical rod-assembly as shown in (e).
The Fourier amplitudes are normalized by the values at $C_{\rm {rod}}=0$ 
(denoted by the superscript $*$).
The error bars are displayed only for the canonical simulations.
The errors in REMD are smaller than the thickness of the lines.
}
\label{fig:rq}
\end{figure}

\begin{figure}
\includegraphics{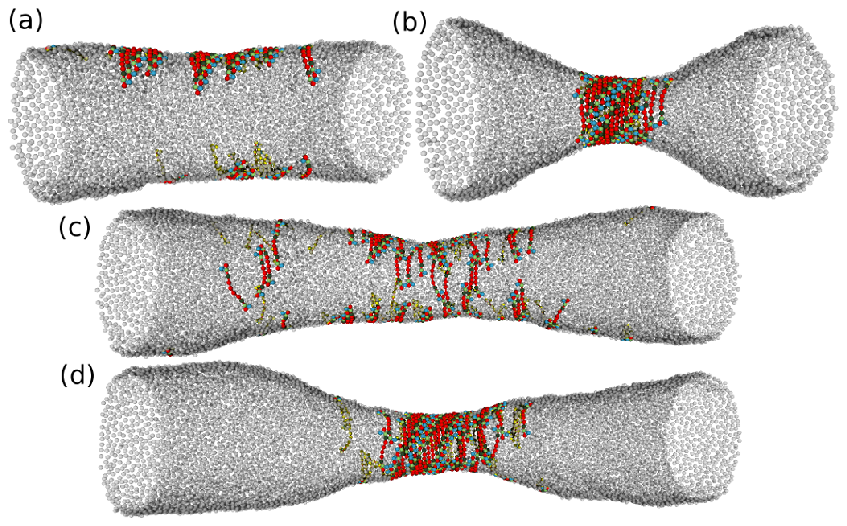}
\includegraphics{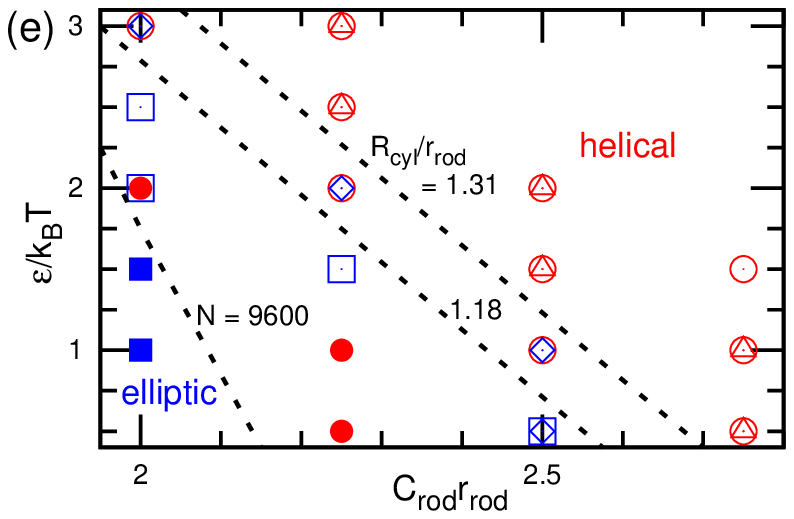}
\caption{
Membrane tube deformation induced by chiral protein rods 
with direct attraction between rods.
(a)--(d) Snapshots of membrane tubes at $R_{\rm cyl}/r_{\rm {rod}}=1.31$.
(a) $\varepsilon/k_{\rm B}T=2$ and (b) $2.5$ at $C_{\rm {rod}}R_{\rm {cyl}}=2.25$ and $N=4800$.
(c) $\varepsilon/k_{\rm B}T=1.5$ and (d) $2$ at $C_{\rm {rod}}R_{\rm {cyl}}=2$ and $N=9600$.
(e) Phase diagram for rod assembly of the elliptical tube and helical-cylinder-shape.
The open circles (triangles) and squares (diamonds) represent the elliptical and helical tubes
at $R_{\rm cyl}/r_{\rm {rod}}=1.18$ ($1.31$) for $N=4800$, respectively.
The closed circles and squares are for $N=9600$ at both $R_{\rm cyl}/r_{\rm rod}=1.18$ and $1.31$.
The dashed lines indicate the phase boundary.
}
\label{fig:att}
\end{figure}

Figure~\ref{fig:rq} shows the rod assembly via membrane-mediated interactions 
for the achiral and chiral rods shown in Figs.~\ref{fig:model}(a) and (b), respectively.
For low rod curvatures of $C_{\rm rod}$, there is no qualitative difference observed between chiral and achiral rods.
As the rod curvature $C_{\rm rod}$ increases, the orientation of the rods rotates from the longitudinal ($z$) direction 
to the azimuthal ($\theta$) direction [Figs.~\ref{fig:rq}(a) and (b)].
With a further increase in $C_{\rm rod}$, the membrane deforms into an elliptical tube and the rods accumulate at the edges of the ellipse
 [Fig.~\ref{fig:rq}(c)].
An even further increase in $C_{\rm rod}$ induces the rod assembly also in the $z$ direction,
and the rest of the tube forms a more rounded shape [Figs.~\ref{fig:rq}(d) and (f)].
The amplitudes of the Fourier modes of the rod density and membrane shape are shown in Figs.~\ref{fig:rq}(g) and (h), respectively:
$r_{qz}= (1/N)\sum_i r_{{\rm 2D},i} \exp(-2\pi z_i {\rm i}/L_z)$ and $r_{q\theta}= (1/N)\sum_i r_{{\rm 2D},i} \exp(-2 \theta_i {\rm i})$
where $r_{{\rm 2D},i}^2 = x_i^2+ y_i^2$ and $\theta_i=\tan^{-1}(x_i/y_i)$.
The subscripts $qz$ and $q\theta$ represent the lowest Fourier modes along the $z$ and $\theta$ directions, respectively.
Both amplitudes of the membrane shape $r_{q\theta}$ and rod density $n_{q\theta}$ 
along the  $\theta$ direction increase together,  
and the amplitudes of $r_{qz}$ and rod density $n_{qz}$ along the $z$ direction increase later.
Thus, the membrane deformation and rod assembly independently occur in the longitudinal and azimuthal directions.
The details of these shape changes of the achiral rods are described in our previous papers~\cite{nogu14,nogu15b,nogu16a}.
Overall, the transition points of the chiral rods are greater than those of the achiral rods.
However, this is not due to the chirality itself but rather to the larger excluded volume interactions of the chiral rods.

By contrast, at high curvatures of $C_{\rm rod}$, it is found that the chiral rods assemble into helical strips and deform the membrane into a cylindrical shape, as shown in Fig.~\ref{fig:rq}(e).
This helical assembly coexists with the elliptical assembly for the short and wide membrane tube ($N=4800$ and $R_{\rm cyl}/r_{\rm rod}=1.31$),
but the  elliptical assembly becomes unstable and spontaneously transforms into the helical assembly 
for the longer or narrower tubes ($N=9600$ or $N=4800$ and $R_{\rm cyl}/r_{\rm rod}=1.18$), as shown in Movie 1 
in the Supplemental Material.
This shape transition causes the rods to be packed in the helical assembly so that the membrane becomes axisymmetric and narrow [Figs.~\ref{fig:rq}(g) and (h)].
In the elliptical tubes, the chiral rods form oligomers with a helical-strip shape [Fig.~\ref{fig:rq}(d)]
but the oligomer size is restricted by the elliptic edge, 
since the large oligomers stick out from the highly curved region of the edge.
By removing the flat region of the elliptical tube, the rods can form a single large assembly on the circular tube.
However, the achiral rods do not form this helical assembly;
even if the helical structure is set as an initial conformation, it quickly transforms back to the elliptical tube.
Thus, the chirality appears to be essential to form this circular tube formation.

Although it has been generally accepted that the direct attractive interactions between specific sites of the BAR domains are essential for the helical assembly, the results of the present simulation revealed that these interactions are in fact not necessary.
Instead, the direct attractions between the protein segments largely promote the assembly.
To further clarify these attraction effects,
the excluded potential $U_{\rm rep}$ is replaced
by the Lennard-Jones (LJ) potential ($U_{\rm LJ}=\sum 4\varepsilon[(\sigma/r_{ij})^{12}-(\sigma/r_{ij})^6)]$) 
between the second and third particles from both rod ends
[Fig.~\ref{fig:model}(c)]. 
The rods gain this side-to-side attraction when they assemble into a helical strip.
Thus, the helical-cylinder-shaped rod-assembly can be stabilized by this attraction [Figs.~\ref{fig:att}(b) and (d) and Movie 2 in the Supplemental Material].
On the other hand, in the elliptical membrane tube, the rods can form only oligomers  as observed for the rods without the direct attractions [Fig.~\ref{fig:att}(a)].
Therefore, this direct side-to-side attraction enhances the formation of the helical assembly, as shown in the phase diagram of Fig.~\ref{fig:att}(e). The greater attraction then induces the helical formation at lower curvatures of $C_{\rm rod}$.
This transition point slightly depends on the average tube radius $R_{\rm cyl}$ for the short tube ($N=4800$),
while no such radius dependence is obtained for the long tube ($N=9600$).
Thus, the radius $R_{\rm cyl}$ has only slight effects on the rod assembly.

\begin{figure}
\includegraphics{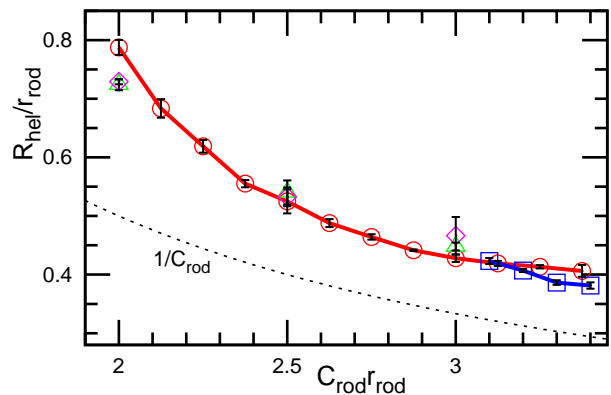}
\caption{
Radius $R_{\rm hyl}$ of the rod assembly of the helical-cylinder shape
for the chiral rods with ($\circ, \triangle, \diamond$) and without ($\Box$) direct side-to-side attractive interactions.
The circles and squares represent $R_{\rm hyl}$ of the rods at $R_{\rm cyl}/r_{\rm rod}=1.18$ and $N=4800$.
The triangles and diamonds are for $R_{\rm cyl}/r_{\rm rod}=1.18$ and $1.31$, respectively, at $N=9600$.
For the side-to-side attraction, $\varepsilon/k_{\rm B}T=3$ is used.
The dashed line shows the curvature radius of the rods $1/C_{\rm rod}$.
}
\label{fig:rcyl}
\end{figure}

The radius $R_{\rm hyl}$ of the helical assembly is a monotonically decreasing function of $C_{\rm rod}$
and exhibits little dependence on the other parameters as shown in Fig.~\ref{fig:rcyl}.
The radius $R_{\rm hyl}$ is calculated
by the least-squares fit for the slice of the middle region with a width of $0.4r_{\rm rod}$
as $R_{\rm hyl} = (1/N_{\rm sl})\sum_i |{\bf r}_{{\rm 2D},i}-{\bf r}_{\rm 2D,g}|$,
where $N_{\rm sl}$ is the number of the fit particles
and ${\bf r}_{\rm 2D,g}$ is the center of the mass projected on the $xy$ plane.
Since the rods are not completely rigid,
$R_{\rm hyl}$ is slightly greater than the preferred radius of the rod curvature $1/C_{\rm rod}$.
For the elliptical tube, the mean tube radius can be largely varied,
since the flat region of the ellipse can be increased by the addition of protein-unbound membranes.
In contrast, the tube radius of the helical rod-assembly is uniquely determined by the protein.
This finding shows good agreement with experimental evidence that each type of BAR protein
typically generates a constant radius of the membrane tubules~\cite{itoh06,masu10,zhao11,mim12a,simu15a}.

Here, we consider a simple achiral side-to-side attractive interaction between protein rods
and the chirality is only generated by the excluded volume of the right-handed hook particles 
in order to clarify the general aspects of the chirality effects.
BAR proteins often have multiple interaction sites, and
some of these proteins also exhibit tip-to-tip interactions \cite{fros08,mcdo15}, which
likely connects two helical strips to stabilize the assembly.
The F-BAR protein, Pascin, induces membrane tubes over a wide diameter range
and is considered to have two types of the assembly structures \cite{wang09}.
Determining the effects of such specific interactions on the membrane tubulation would therefore be 
an interesting problem for the further investigations.

In summary, the present coarse-grained simulation demonstrates that the chirality of proteins
induces helical rod assembly on a cylindrical membrane of a constant radius.
The side-to-side direct attraction between proteins stabilizes this assembly.
Although 
 the proteins can still induce membrane tubules without the chirality,  
the shape is elliptical and the radius is not constant.
The helical interaction induces the protein assembly not only in the side-to-side direction but also in the tip-to-tip direction
leading to formation of a circular tube.
This scenario seems to be more efficient than the case, in which different assembly mechanisms 
are employed for the side and tip directions.
This type of the helical interaction may also play an important role in other protein assemblies, such as those occurring 
on biomembranes including dynamins~\cite{anto16} and ESCRT~\cite{raib09}.

\begin{acknowledgments}
The RMD simulations were
carried out by HPE SGI 8600
at ISSP Supercomputer Center, University of Tokyo. 
This work was supported by JSPS KAKENHI Grant Number JP17K05607.
\end{acknowledgments}

\end{document}